\documentclass{ccjnl}
\title{Limited Feedback in RIS-Assisted Wireless Communications: Use Cases, Challenges, and Future Directions}
\author{Weicong Chen\inst{1}, Jiajia Guo\inst{1}, Yiming Cui\inst{1}, Xiao Li\inst{1}, Shi Jin\inst{1,*}\corinfo{jinshi@seu.edu.cn}} 
\receiveddate{XX XX, 202X}
\reviseddate{XX XX, 202X}
\Editor{XX XX}

\address[1]{The National Mobile Communications Research Laboratory, Southeast University, Nanjing, 210096, P. R. China}% anonymous
\begin{document}
\maketitle

\begin{abstract}
Channel state information (CSI) is essential to unlock the potential of reconfigurable intelligent surfaces (RISs) in wireless communication systems. Since massive RIS elements are typically implemented without baseband signal processing capabilities, limited CSI feedback is necessary when designing the reflection/refraction coefficients of the RIS. In this article, the unique RIS-assisted channel features, such as the RIS position-dependent channel fluctuation, the ultra-high dimensional sub-channel matrix, and the structured sparsity, are distilled from recent advances in limited feedback and used as guidelines for designing feedback schemes. We begin by illustrating the use cases and the corresponding challenges associated with RIS feedback. We then discuss how to leverage techniques such as channel customization, structured-sparsity, autoencoders, and others to reduce feedback overhead and complexity when devising feedback schemes. Finally, we identify potential research directions by considering the unresolved challenges, the new RIS architecture, and the integration with multi-modal information and artificial intelligence.
\end{abstract}

\section{Introduction}

Having merged from metasurfaces, reconfigurable intelligent surfaces (RISs) are envisioned to realize a smart radio environment by fine-tuning electromagnetic waves in the wireless channel \cite{smart-radio}. With the ability to customize wireless channels, RIS offers one more degree of freedom to optimize wireless communication systems. \textcolor{black}{The potential of RIS has recently been exploited in various wireless applications, such as millimeter wave communications \cite{CC_mmwave}, integrated sensing and communication \cite{RIS-ISAC}, unmanned aerial vehicle communications \cite{RIS-Veh}, simultaneous wireless information and power transfer \cite{Sun3}, mobile edge computing \cite{RIS-MEC}, anti-jamming communications \cite{Sun1,Sun2}, and physical layer security \cite{TFIS}. \textcolor{black}{The extensive research on the applications of RIS has spurred real-world measurement validation campaigns \cite{JSang,5G_Liu,5G_mmwave}. In 5G commercial sub-6 GHz networks assisted by RIS, a maximum gain of 4.03 dB in the reference signal received power was achieved as reported in \cite{JSang}. Additionally, a 275$\%$ improvement in the downlink data rate was demonstrated in \cite{5G_Liu}. Furthermore, field trials for RIS-assisted commercial millimeter wave communications showed that RIS can provide more than 16 dB of gain for the beam of interest, as indicated in \cite{5G_mmwave}.} In addition to theoretical analysis and real-world measurements, RIS standardization activities are under intense debate by standards organizations \cite{Q.Wu-standard}.} 

However, the success of various RIS applications is heavily dependent on the availability of channel state information (CSI), which is essential for configuring the surfaces. Acquiring CSI in RIS-assisted wireless systems presents unique challenges compared to systems without RISs, as RISs are typically composed of a multitude of passive elements without signal processing capabilities. Deviating from the traditional sequential process of channel estimation followed by RIS configuration, the authors in \cite{JSAC_CC} proposed an innovative approach that integrates CSI acquisition and RIS configuration, leveraging the channel customization capabilities of the RIS. Channel estimation becomes necessary to obtain CSI and has been the subject of extensive research \cite{Swindlehurst}. In frequency division duplex (FDD) systems, the downlink and uplink channels are not reciprocal, necessitating the feedback of estimated downlink CSI to the base station (BS) for downstream transmission design.

Due to the finite capacity of feedback links, information fed back to the BS must be limited \cite{Nihar}. In wireless systems without RIS, limited feedback has been widely studied \cite{DJLove}. Limited feedback can be divided into two main categories based on the content to be fed back. The first category aims to feed back the complete CSI to enable the BS to reconstruct the downlink channel matrix. The limited channel multipath parameters and the compressed channel matrix are fed back for channel reconstruction subject to the feedback capacity. The second category focuses on feeding back the dominated channel component using the predefined codebook shared by the BS and the user. By feeding back the precoder matrix indicator (PMI), the BS can identify the precoder. In current wireless systems, limited feedback is designed for the end-to-end channel between the BS and the user.

\begin{figure*}[h]
	\centering
	\includegraphics[width=0.8\textwidth]{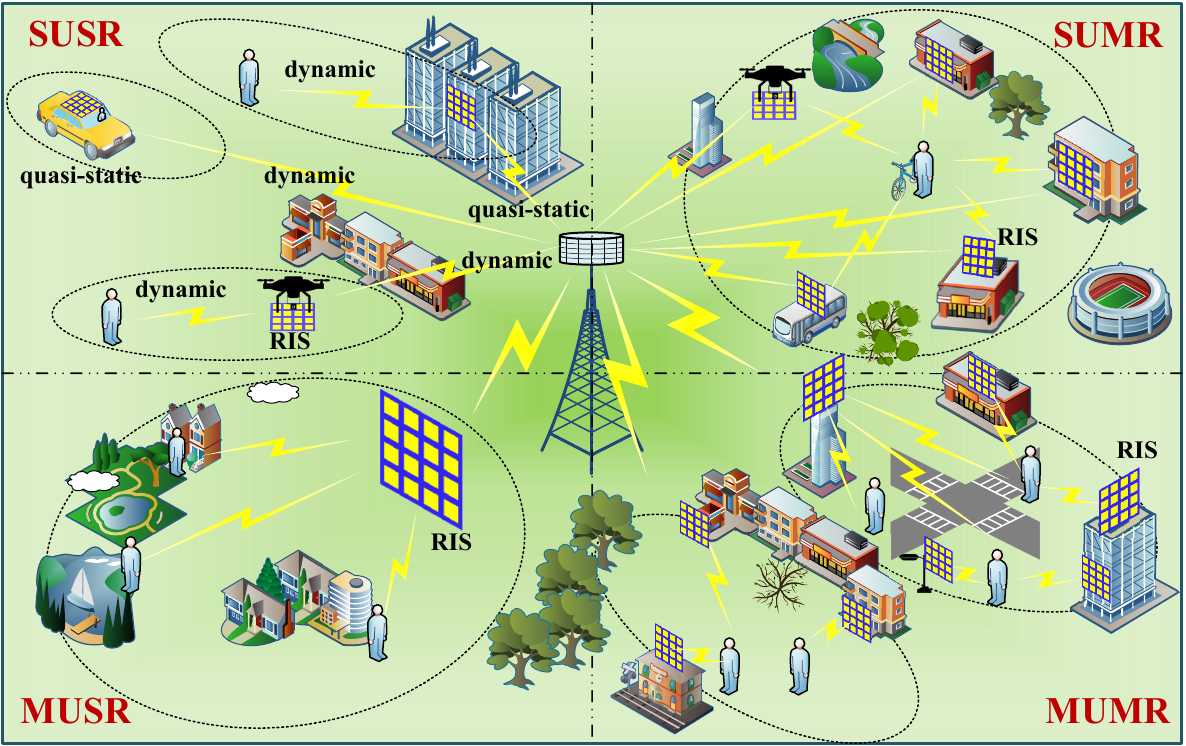}
	\caption{Channel characteristics in SUSR, SUMR, MUSR, and MUMR scenarios.}
	\label{Fig.four-cases} %% label for entire figure
\end{figure*}

To realize a smart radio environment with more degrees of control freedom, it becomes necessary to deploy multiple RISs, each equipped with a multitude of elements, between the BS and the user. However, the abundance of propagation paths and high-dimensional channel matrices create an overwhelming overhead for feedback. Furthermore, in addition to the conventional codebook for the PMI feedback, another codebook proportional in size to the number of RIS elements must be designed to feed back the RIS configuration when adopting a codebook-based feedback scheme \cite{JAn}. As a result, existing limited feedback schemes cannot be simply extended; they must be reevaluated to accommodate the novel channel features introduced by the RIS. 

\textcolor{black}{In RIS-assisted wireless systems, \cite{FB-1} and \cite{FB-2} briefly touch on feedback, and several feedback schemes \cite{ Shen-tcom2021}--[34] have been developed. Leveraging the single-structured sparsity of the BS-RIS-UE cascaded channel, \cite{ Shen-tcom2021} introduced a dimension-reduced feedback scheme to alleviate the substantial feedback burden. Building on this foundation, the authors in \cite{ Shi-CL2022} capitalized on the triple-structured sparsity of the beamspace cascaded channel, further diminishing the feedback overhead. Different from \cite{ Shen-tcom2021} and \cite{ Shi-CL2022} which reduces the number of channel parameters, \cite{ shin-access} proposed a compressive sensing-based feedback scheme that compresses the CSI with a small number of channel vectors. Utilizing the compression and recovery capabilities of deep learning (DL), the convolution network, Quan-Transformer, and autoencoder were employed to reduce the feedback overhead in \cite{ ZhangjiePeng}, \cite{ xie2022quan}, and \cite{ cui2024efficient}, respectively. Under limited feedback, \cite{Chen-twc2022}--[29] aimed to improve the CSI recovery accuracy at the BS. A novel cascaded codebook along with an adaptive feedback bit allocation algorithm was proposed in \cite{Chen-twc2022} to flexibly quantize the diverse line-of-sight (LoS) and non-LoS (NLoS) path gains in the BS/user-RIS channel, minimizing the quantization error when the number of feedback bits is fixed. Considering that both CSI estimation and feedback introduce errors for the CSI recovery process, the authors in \cite{ Joint-CECF} proposed to directly incorporate preliminary channel estimation with errors into the compressed feedback process, enabling the joint optimization of CSI estimation and feedback. This joint design outperforms some benchmark methods in terms of CSI recovery accuracy. To enhance CSI reconstruction accuracy through DL, \cite{ Yu-CL2022} removed the batch normalization layers and introduced a denoising module to solve the problems of mismatched distribution and vanishing gradient, and \cite{ RISMCNet} developed a novel DL-based solution by utilizing the two-timescale feature and the sparsity of the RIS-assisted channel. The impact of limited feedback on the transmission performance was investigated in \cite{ MUMR-Cheng}--[34]. Specifically, \cite{ MUMR-Cheng} found that spectral efficiency loss from finite-resolution codebooks for RIS is independent of transmit power in high signal-to-noise ratio regimes, whereas loss due to limited CSI feedback increases with the transmit power. The authors in \cite{ RIS-RVQ} obtained an upper bound for the rate loss in the Rayleigh fading channel as a function of the number of feedback bits and the size of RIS. Utilizing RISs to reshape a sparse channel, the joint transceiver and RIS design with quantized CSI can achieve satisfactory performance compared with the optimal transceiver with full CSI \cite{ Chen-twc-CC}. In \cite{ Guo-tvt}, a DL-based two-timescale CSI feedback was proposed to directly produce the beamforming vector and RIS phase shifts according to the received feedback information, outperforming conventional algorithms in terms of achievable rate. Ref. \cite{ Kim-twc2022} proposed an adaptive codebook-based limited feedback protocol, improving the data rate performance by taking into account the practical RIS reflection behavior.}

\textcolor{black}{Although existing studies have explored limited feedback in RIS-assisted systems by reducing feedback overhead, enhancing CSI recovery accuracy, and improving transmission performance, the unique characteristics introduced by RIS have not been fully summarized and revealed. This paper aims to address this gap for potential standardizations by demonstrating the design of a limited feedback scheme in FDD systems, considering the specific characteristics of RIS-assisted channels. The RIS-associated limited feedback is divided into two primary categories: channel reconstruction and RIS configuration. We distill the notable RIS-introduced channel features, such as the RIS position-dependent channel fluctuation, the ultra-high dimensional sub-channel matrix, the time correlation, and the structured sparsity, in various user--RIS topologies to guide the channel reconstruction. Given that the RIS is introduced as a new node to be configured in wireless networks, new challenges, such as the interplay between active and passive beamforming, the extensive configuration parameters, and the angle-dependent phase shifts, are discussed for the design of RIS configuration feedback. Based on these insights, we explore the potential limited feedback strategies for channel reconstruction and RIS configuration in detail. Finally, prospective research directions in this area are highlighted.}

\section{Use Cases of RIS-Associated Limited Feedback}\label{sec:2}

Functioning as a new type of system node, RIS enriches the application scenarios for limited feedback. The mobility, control mechanism, deployment scenarios, and other relevant aspects of RIS introduce additional features and challenges in the design of feedback schemes. On one hand, the introduction of RIS establishes an artificial link comprising two sub-channels: the BS--RIS channel and the RIS--user channel. Therefore, when employing limited feedback to reconstruct the downlink channel, these additional sub-channels must be taken into consideration. On the other hand, widely adopted passive RISs require control information from the BS and/or user due to their lack of signal processing capabilities. Utilizing limited feedback to transmit control information for configuring the RIS represents a novel scenario compared to wireless systems without RISs. We categorize the use cases of RIS-associated limited feedback into two: channel reconstruction and RIS configuration. In the following subsections, we detail the pertinent features and challenges that stem from the two use cases.

\subsection{Channel Reconstruction}\label{sec:2.1}
Channel reconstruction at the BS side empowers the BS with complete channel knowledge, enabling full control of the entire cellular network. However, due to the limited feedback capacity, users can only transmit a finite set of channel parameters back to the BS for channel matrix reconstruction. \textcolor{black}{Establishing cascaded paths through RISs to aid communication also introduces a greater number of channel parameters, posing a significant challenge to CSI feedback.} These parameters exhibit distinct characteristics compared to those in conventional systems, primarily owing to their dependence on various user-RIS topologies. \textcolor{black}{Considering the varying numbers of users and RISs, we categorize these user-RIS topologies into four scenarios with increasing system complexity, as illustrated in Fig. \ref{Fig.four-cases}.}  

\subsubsection{Single-User Single-RIS (SUSR) Scenarios}
In this basic scenario, the cascaded channel might be intricate due to varying deployment states of the RIS. When the RIS is installed on tall buildings, the BS--RIS channel remains relatively static with a dominant LoS path, while the RIS--user channel is dynamic with multiple propagation paths. Conversely, when the RIS is mounted on a vehicle to provide communication services to passengers inside, the BS--RIS channel changes rapidly over time, and the RIS--user channel stays more constant. By placing the RIS on an unmanned aerial vehicle, both the BS--RIS and RIS--user channels can be dynamic with strong LoS paths. These diverse channel features suggest that traditional feedback methods using codebooks on a single timescale might not be directly applicable \cite{Chen-twc2022}.

\subsubsection{Single-User Multiple-RIS (SUMR) Scenarios}
\textcolor{black}{Beyond the channel characteristics of SUSR scenarios, introducing multiple RISs to assist a single user establishes a multitude of cascaded paths.} \textcolor{black}{Due to the multiplicative path loss introduced by the RIS, multi-reflection channels between RISs are negligible compared to the single-reflection channel. Given the RIS's large element count and the radio environment's abundant  paths in the sub-6 GHz band, these high-dimensional single-reflection channels can still lead to overwhelming feedback overhead.} Note that the maximum link gain of the cascaded channel is determined not only by the RIS scale but also by its placement, as the cascaded path loss is tied to the distances between the RIS and the BS/user. Hence, it is crucial to allocate feedback resources based on the RIS's position and size.

\subsubsection{Multiple-User Single-RIS (MUSR) Scenarios}
\textcolor{black}{Similarly evolving from the SUSR scenario, but differing from the SUMR scenario, when a single RIS serves multiple users, a new channel characteristic emerges where different users share the same BS-RIS channel.} By using this shared channel to create a combined feedback scheme for all users, the overall feedback overhead can be minimized. For instance, while the individual RIS--user channel should be relayed back separately by each user, the shared BS--RIS channel could be sent by just one user. Once the RIS-cascaded channels are accessible, the structured sparsity of the cascaded channel matrix can be used to lessen feedback overhead. This structured sparsity is present in the beamspace channel since the angle of departure (AoD) at the BS, the angle of arrival (AoA) offset at the RIS, and the cascaded path gain ratios remain consistent across users \cite{Shen-tcom2021} \cite{Shi-CL2022}.

\subsubsection{Multiple-User Multiple-RIS (MUMR) Scenarios}
\textcolor{black}{In the most general MUMR scenarios, beyond the channel features already discussed, the RIS's relative position to the user dictates the path loss, and different cascaded channels hold varying significance for each user.} As a result, both RIS and user associations should be taken into account \cite{MUMR-Cheng} to maximize the total gain from these distributed RISs. Since the associated RISs are generally set to boost a particular user's transmission link, these RISs create negligible cascaded channels for others. Therefore, each user might only need to feedback its own associated cascaded channel without omitting essential channel data. Put differently, leveraging channel features to devise joint user feedback and RIS setup can help decrease the feedback overhead.

\subsection{RIS Configuration}
Compared to sending back the channel parameters to the BS for channel reconstruction, directly sending RIS configuration instructions through a codebook-based scheme is a more efficient approach. \textcolor{black}{Given that the RIS is an additional node in wireless networks, its resulting features should be taken into account when dealing with limited feedback. Distinct from Sec. \ref{sec:2.1}, which organizes channel reconstruction in a progressive manner, this section discusses the challenges of RIS configuration from three parallel  aspects: the balance between active beamforming at the BS and passive beamforming at the RIS, the large number of RIS configuration parameters, and the angle-dependent phase shifts.}

\subsubsection{Active and Passive Beamforming}
Given that the RIS provides a controllable link, it becomes vital to account for active beamforming at the BS alongside passive beamforming at the RIS. In scenarios where the RIS is operated by the BS via wired links, feedback is exclusive to the BS--user channel. Therefore, striking the right balance in feedback overhead for both active and passive beamforming is key to achieving optimal performance within the constraints of feedback capacity. In instances where the RIS is not linked through a wire, the feedback standard necessitates a redesign. Such situations involve creating an extra wireless feedback connection, either from the user or the BS to the RIS, for the relay of configuration details.

\subsubsection{Massive Configuration Parameters}\label{sec:2-B-2}
In traditional communication systems, the precoder for the downlink is determined by the feedback of the PMI, with codebooks designed to accommodate systems with dozens of antennas. \textcolor{black}{However, given that the RIS possesses a vast array of elements to counteract significant path losses, there arises a need for a more expansive codebook for its proper passive beamforming configuration.} Traditional codebooks, such as the discrete Fourier transform (DFT) codebook \cite{DFT} and the random vector quantization (RVQ) codebook \cite{RVQ} \cite{RIS-RVQ}, become less effective due to the pronounced feedback overhead they introduce. To efficiently configure the RIS within acceptable feedback constraints, a reevaluation of both codebook structures and feedback strategies is imperative.

\subsubsection{Angle-dependent Phase Shift}
Operating as an intermediary, the RIS directs electromagnetic (EM) waves from unmanaged to preferred directions by applying phase shifts on the incoming wavefront. The phase shift of the RIS is influenced both by the capacitance of the adjustable diode and by the incident angle of EM waves, a result of spatial dispersion \cite{angle-d}. As such, when the desired phase shifts are dispatched to the RIS without the corresponding incident angles, it becomes unfeasible to identify the correct capacitances needed to set the RIS elements to achieve those shifts. This underscores the importance of constructing the codebook based on capacitances rather than merely detailing phase shifts.

%\begin{figure*}[!h]
%	\centering
%	\includegraphics[width=1\textwidth]{THREE.pdf}
%	\caption{Feedback mechanism: (a) when the RIS is controlled by the BS via wire link, there exists feedback link from the user to the BS; (b) without wire link to control the RIS, additional feedback link from the BS to the RIS is required; (c) without wire link to control the RIS, additional feedback link from the user to the RIS is required.}
%	\label{Fig.THREE} %% label for entire figure
%\end{figure*}

%%%%%%%%%%%%%%%%%%%%%%%%%%%%%%%%%%%%%%%%%%%%%%%%%%%%%%%%%%%%%%%%%%%%%%%%%%%%%%%%%%%%%%%%%%%%%%

\section{Limited Feedback for Channel Reconstruction}\label{sec:3}

In this section, we present potential limited feedback schemes for channel matrix reconstruction, considering various user-RIS topologies. These schemes encompass codebook-based feedback, channel customization-based feedback, DL-based feedback, and structured sparsity-based feedback. From these methods, \textbf{we highlight how the channel features introduced by the RIS can be utilized to improve feedback performance while reducing overhead}.

\subsection{Codebook-based Feedback}
\begin{figure}[!t]
	\centering
	\includegraphics[width=0.5\textwidth]{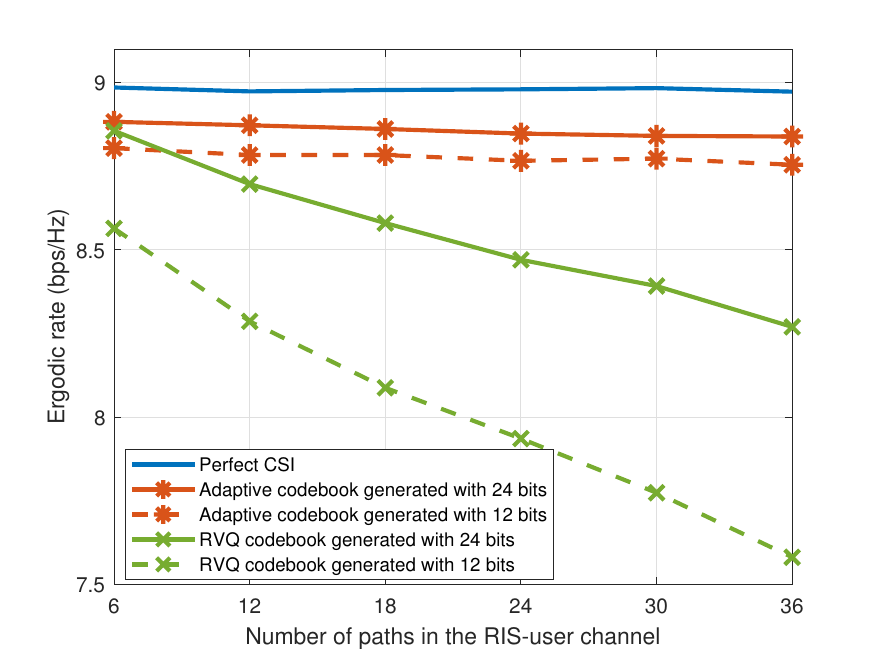}
	\caption{Performance of the adaptive codebook \cite{Chen-twc2022} and the RVQ codebook.}
	\label{Fig.bit_par} %% label for entire figure
\end{figure}

The segmentation of the cascaded channel into two sub-channels by the RIS, with their potential differences in distribution based on the RIS's deployment state, implies that traditional single-structure codebooks may not be directly applicable for feedback schemes. Recognizing this, \cite{Chen-twc2022} proposed an adaptive cascaded codebook tailored for SUSR systems. Designed to provide feedback for cascaded path gains, this codebook synthesizes a codeword from two sub-codewords, each corresponding to one of the two sub-channels. Both LoS and NLoS components are encapsulated within each sub-codeword, ensuring that \textbf{the codeword structure is consistent with the cascaded path gain's characteristics}.

With a fixed number of feedback bits, the dynamic allocation of these constrained bits according to the channel distribution facilitates the creation of cascaded codewords. \textcolor{black}{Because the adaptive cascaded codeword captures the importance of elements in the path gain vector, it outperforms the conventional RVQ codeword in terms of ergodic rate when the maximum ratio combining is designed with the CSI reconstructed by the optimal codeword, as depicted in Fig. \ref{Fig.bit_par}. Given that the dimension of path gain vector increases with more paths, the quantization accuracy of the RVQ codeword decreases, leading a performance degradation. In contrast, the adaptive codeword can dynamically allocate feedback bits based on the importance of each path gain, offering better robustness.}

\subsection{Deep Learning-based Feedback}
Deep learning has shown considerable promise in CSI feedback challenges \cite{JiaOverview}. Notably, DL-enabled autoencoder-based feedback enhancement was highlighted as a representative use case in the 3rd Generation Partnership Project (3GPP) Release 18 \cite{3GPP}. By training an autoencoder network with numerous channel samples, the network can automatically extract sophisticated characteristics inherent in the BS--RIS--UE channel to facilitate efficient CSI compression and reconstruction. Different from the conventional DL-based CSI feedback in massive MIMO systems, novel neural architectures such as channel attention \cite{ZhangjiePeng} and multi-head attention \cite{xie2022quan} are preferred when designing the autoencoder, catering to the larger channel dimension and more complicated channel structure of the BS--RIS--UE channel.

To further reduce the feedback overhead, time correlation among channels in adjacent time intervals can be exploited with delicate neural architecture design. Long short-term memory (LSTM) is a prevalent module in processing time sequence data. Leveraging a low-complexity variant named convolutional LSTM, \cite{ZhangjiePeng} achieves significant performance gain compared to the methods without time correlation extraction, especially when feedback length is severely limited. While \cite{ZhangjiePeng} directly extracts time correlation from cascaded channel in a black-box way, \cite{cui2024efficient} advocates discriminating different time correlation characteristics between BS--RIS channel and RIS--UE channel, further improving the feedback efficiency.

%\textcolor{red}{\cite{cui2024feedback} and \cite{ZhangjiePeng}}

\subsection{Channel Customization-based Feedback}
\begin{figure}[!t]
	\centering
	\includegraphics[width=0.5\textwidth]{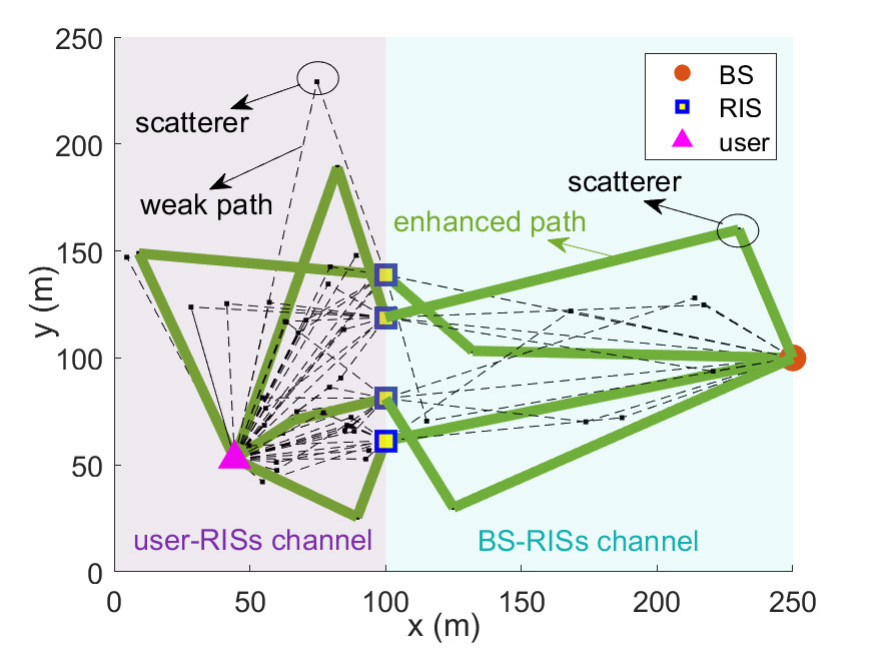}
	\caption{Channel customization-based feedback that reshapes the complex composite channel to be sparse \cite{Chen-twc-CC}.}
	\label{Fig.CC} %% label for entire figure
\end{figure}
\begin{figure*}[t]
	\centering
	\subfloat[The spatial domain channel]{
		\includegraphics[width=0.305\textwidth]{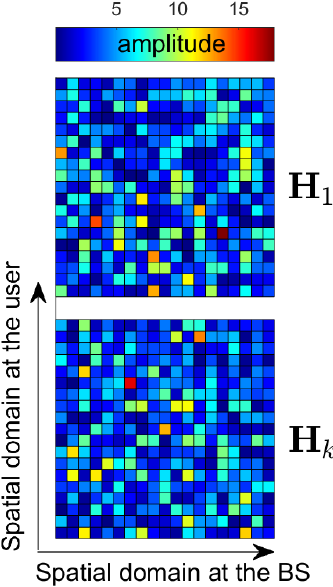}}
	\subfloat[The hybrid domain channel]{
		\includegraphics[width=0.3\textwidth]{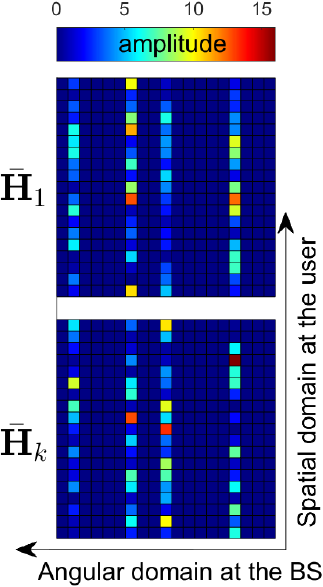}}
	\caption{When the RIS-cascaded spatial domain channel ${\bf H}_k$, whose element denotes the channel responses of each antenna-RIS unit pair in space, is projected into hybrid spatial and angular domains channel $\bar{\bf H}_k$ using the DFT matrix,  different users share the same indices of the non-zero columns, which is defined as single-structured sparsity \cite{Shen-tcom2021}.}
	\label{Fig.single} %% label for entire figure
\end{figure*}
In SUMR MIMO systems, the incorporation of multiple RISs results in a multitude of channel paths, incurring a considerable feedback overhead for channel reconstruction. This poses challenges in the joint design of an optimal singular value decomposition (SVD) transceiver with the RIS's phase shifts,  since reconstructing the matrices of the downlink channel component is essential.

To reconstruct the downlink composite channel while simultaneously deploying an SVD transceiver with minimal feedback overhead and computational complexity, \cite{Chen-twc-CC} put forth \textbf{a channel customization approach that transforms the intricate composite channel into a sparse structure}, as visualized in Fig. \ref{Fig.CC}. Within this framework, users relay to the BS the direction parameters of specific orthogonal paths, enabling the BS to configure the relevant RISs accordingly. By accentuating these orthogonal paths and diminishing other paths, the channel undergoes a transformation where the rich scattering channel is predominantly governed by a handful of orthogonal pathways. This paves the way for approximating the composite channel through SVD. To further mitigate feedback overhead during the implementation of water-filling power allocation for the SVD transceiver, \cite{Chen-twc-CC} adjusts the RISs based on the path loss associated with them. This adjustment ensures the singular values of the customized channel are relatively congruent in magnitude.

\subsection{Structured Sparsity-based Feedback}

The feedback strategies proposed in \cite{Chen-twc2022} and \cite{Chen-twc-CC} possess the potential to be extrapolated to multi-user contexts, given the premise that the parametric channel can be discerned by the user. By exploiting the sparse nature of the millimeter wave channel, compressive sensing based feedback schemes can be used to reduce feedback overhead of multiple users \cite{shin-access}. When the user has access to the cascaded channel matrix, the structured sparsity inherent to the cascaded channel matrix is leveraged in \cite{Shen-tcom2021} and \cite{Shi-CL2022} to trim down feedback overhead in MUSR scenarios.

In MUSR contexts, where both the BS and RIS are bounded by sparse scatterers, the BS--RIS channel evinces a distinct sparsity, manifested by a handful of AoDs at the BS. This sparsity becomes even more pronounced when the BS--RIS--user cascaded channel matrix is projected into a hybrid of spatial and angular domains: limited non-zero column vectors emerge. Notably, the count of these non-zero vectors in the angular spectrum corresponds to the AoDs at the BS. This highlights an inherent property: \textbf{the matrices of RIS-cascaded channels, across different users, consistently share non-zero column indices}, visualized in Fig. \ref{Fig.single}. Such consistency is attributed to the universally shared BS--RIS channel amongst users, a phenomenon termed as single-structured sparsity in \cite{Shen-tcom2021}.
\begin{figure*}[!h]
	\centering
	\includegraphics[width=0.8\textwidth]{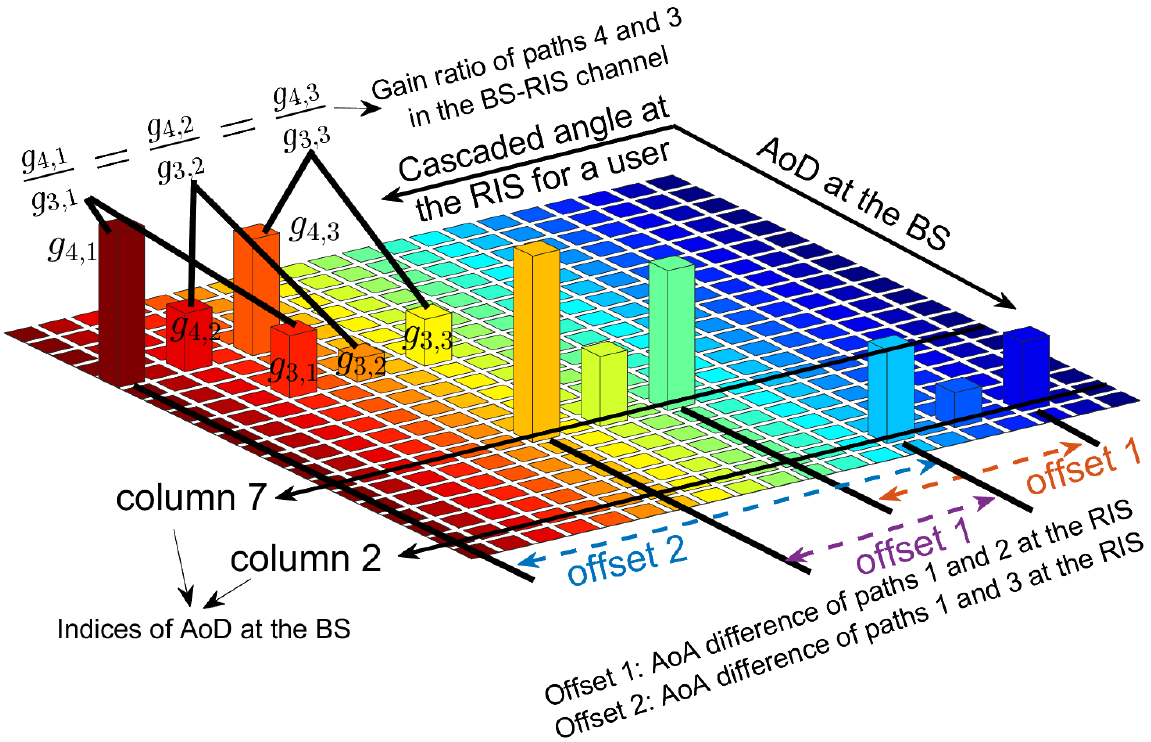}
	\caption{Triple-structured sparsity: in addition to the non-zero column indices, the beamspace cascaded channel exists absolute identity between the non-zero columns except for the location offset and cascaded path gain ratio \cite{Shi-CL2022}. The column height and color-coding of the picture indicate the amplitude of $g_{m,n}$ and the index of the cascaded angle, respectively.} 
	\label{Fig.triple} %% label for entire figure
\end{figure*}
Leveraging this unique sparsity, the feedback associated with the channel can be bifurcated into two distinct classes: feedback concerning non-zero column indices and feedback relating to the non-zero column vectors. Adopting an economical approach to feedback, \cite{Shen-tcom2021} advanced a scheme wherein a singular user relays the non-zero column indices to the BS, effectively acting as a representative for other users. In parallel, each user individually communicates the non-zero column vectors to the BS.

The study in \cite{Shi-CL2022} took a deeper dive into the intricacies of structured sparsity-based feedback. To elucidate, let's envisage a situation where the BS--RIS channel is characterized by four paths, while the channels from RIS to user $k$ have three paths. The differential between the AoD and AoA at the RIS is termed the cascaded angle. The cascaded path gain from the $m$-th path in the BS--RIS channel to the $n$-th path in the RIS--user channel is denoted as $g_{m,n}$. By transforming the RIS-cascaded channel into its beamspace counterpart using the DFT matrices, we identified three distinct sparsity patterns, as depicted in Fig. \ref{Fig.triple}.

Notably, the beamspace cascaded channel, in addition to having non-zero column indices, displays inherent uniformity across non-zero columns. This uniformity showcases noticeable variations in both the location offset of the cascaded angle and cascaded path gain ratio. Such a characteristic can be traced back to paths originating from the BS that split into three cascaded paths via their respective trajectories in the RIS-user channel. Consequently, the non-zero columns of the beamspace cascaded channel primarily reflect the foundational sparse pattern of the RIS-user channel. However, they are modulated by different multiplicative factors and spatial cyclic shifts. Specifically, the multiplicative factors and the spatial cyclic shifts are influenced by the path gains and AoAs at the RIS of the four paths present in the BS-RIS channel.

Because different users share commonalities, like the AoA at the RIS and path gain in the BS--RIS channel, certain parameters tied to offset and path gain ratio are contingent exclusively on the BS--RIS channel and are uniform across users. Harnessing these channel traits, \textbf{shared parameters}---including BS AoDs, location offsets, and path gain ratios---\textbf{can be extracted from the RIS-cascaded channel and selectively relayed back to the BS by designated users.} This exploitation of the structured sparsity offers avenues for further diminishing overhead in multi-user feedback contexts.

\subsection{\textcolor{black}{Comparison of the Presented Methods}}
\textcolor{black}{Both the codebook-based and DL-based feedback are designed for SUSR scenarios, yet their generalization differs. The adaptive codebook proposed in \cite{Chen-twc2022} can cater to different distributions of the propagation paths and is independent of the system antenna number, while the network in DL-based methods might need re-training for a new system setup. For a given system, the DL-based feedback can significantly reduce the feedback overhead by extracting and utilizing the environment information. On the contrary, the effectiveness of the adaptive codebook in reducing feedback overhead diminishes in high-mobility scenarios, as its underlying assumption is that the variation of path direction is much slower compared to the path gain. The channel customization-based feedback is specifically designed to address the rich propagation paths introduced by multi-RIS and also performs well in SUSR scenarios with a large number of paths. The premise of using this method is that among the many propagation paths, there are a few strong ones, and the RIS is equipped with a large number of elements to provide sufficient array gain to sparsely represent the channel. Methods tailored for systems with single-user, while scalable to multi-user scenarios, can not leverage the shared BS--RIS channel to reduce the feedback overhead, as demonstrated in \cite{Shen-tcom2021} and \cite{Shi-CL2022}. Considering the benefits of the structured sparsity are related to the number of paths, the feedback overhead in \cite{Shen-tcom2021} and \cite{Shi-CL2022} increases with the number of paths.}

%%%%%%%%%%%%%%%%%%%%%%%%%%%%%%%%%%%%%%%%%%%%%%%%%%%%%%%%%%%%%%%%%%%%%%%%%%%%%%%%%%%%%%%%%%%%%%
\section{Limited Feedback for RIS Configuration}\label{sec:4}
\begin{figure*}[!h]
	\centering
	\includegraphics[width=0.75\textwidth]{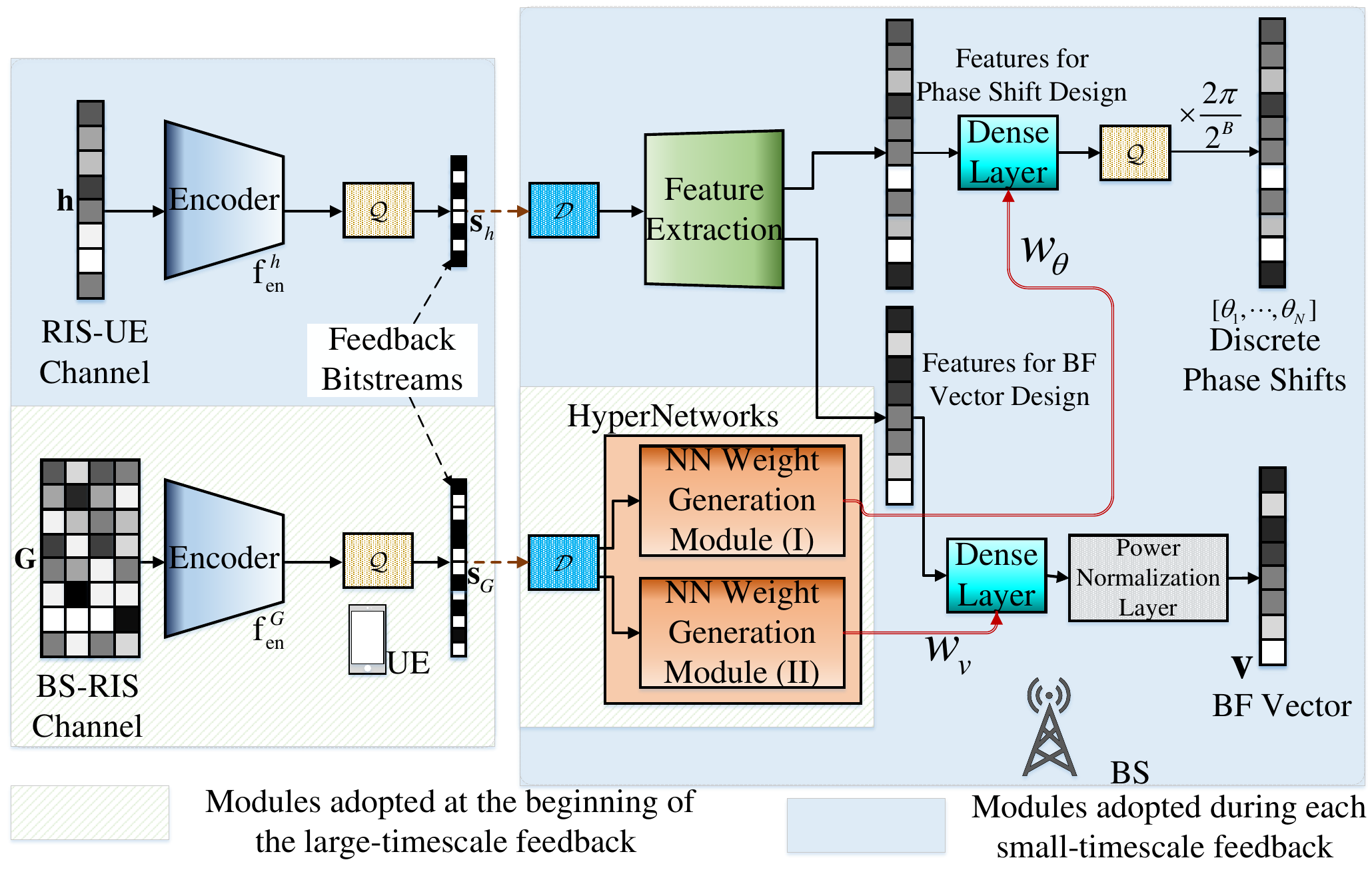}
	\caption{DL-based two-timescale CSI feedback and beamforming framework, RIS-CsiNet \cite{Guo-tvt}.}
	\label{Fig.RIS_CsiNet} %% label for entire figure
\end{figure*}

Compared to channel reconstruction, directly configuring the RIS with a predefined codebook results in reduced feedback overhead, though this comes at the expense of performance degradation. However, designing feedback for the RIS configuration remains challenging due to the large number of RIS elements. As illustrated in Sec. \ref{sec:2-B-2}, the conventional codebook-based schemes, which are efficient in wireless systems without an RIS, may not be as effective when an RIS is introduced. DL-based limited feedback schemes for RIS configuration, aimed at reducing feedback overhead and facilitating adaptive codebook design, have been explored in \cite{Guo-tvt,Kim-twc2022,Yu-CL2022,Yu-arxiv}. In alignment with this DL-based paradigm, we introduce three potential limited feedback schemes for RIS configuration in the subsequent subsections. \textcolor{black}{These three schemes represent distinct feedback protocols: feedback from the user to the BS, from the BS to the RIS, and from the user directly to the RIS.}

\subsection{Two-timescale Feedback from User to BS}

In scenarios where the feedback process is executed from the user to the BS, and the RIS can accurately receive configurations from the BS, a DL-based joint CSI feedback and beamforming framework called RIS-CsiNet \cite{Guo-tvt} was proposed. This framework leverages a two-timescale feedback mechanism that \textbf{exploits the high-dimensional, slow-varying characteristics of the BS--RIS channel and the low-dimensional, fast-varying characteristics of the RIS--user channel}. Specifically, the feedback mechanism operates at two timescales: a large timescale composed of multiple small timescales. \textcolor{black}{At the beginning of each large timescale, the user feeds back the compressed BS--RIS channel information. In contrast, the compressed RIS--user channel information is fed back before data transmission in each small timescale. These compressed channel information will be utilized at the BS to directly produce configuration for the RIS, rather than reconstructing the channel.} The two time-scale solution for feedback was further extended to RIS-assisted multi-carrier systems \cite{RISMCNet}, where the sparsity of the angular-delay domain channel was utilized and an attention-based encoder was applied to reduce the feedback overhead.\par

Fig. \ref{Fig.RIS_CsiNet} illustrates the primary neural network (NN) framework of the RIS-CsiNet in \cite{Guo-tvt}. Two DL-based encoders at the user compress and quantize the BS--RIS and RIS--user channels into bitstreams, respectively. \textcolor{black}{These bitstreams are not employed to recover the CSI, as is the case in Sec. \ref{sec:3}.} Once obtaining the feedback bitstreams at the beginning of each large timescale, the hypernetwork generates the parameters for the NNs that produce the BS beamforming vector and RIS phase shift, thereby avoiding the frequent input of the fixed BS--RIS channel information. \textcolor{black}{At each small timescale, the NNs extract the CSI features from the compressed RIS--user channel information and send these features to the NNs. In this way, the NNs at the BS side directly generate the beamforming vector and RIS phase shift with the compressed BS--RIS and RIS--user channel information, thereby reducing the feedback overhead and computational complexity.}

\subsection{Adaptive RIS Control from BS to RIS}
To achieve effective RIS control under the angle-dependent hardware characteristic, an adaptive codebook-based limited feedback protocol from the BS to the RIS was proposed in \cite{Kim-twc2022}. Unlike existing works that simply design phase shifts for the RIS, the authors \textbf{considered the angle-dependent characteristic and directly designed the capacitance value for each RIS element}. Therefore, the codebook was proposed to be constructed by a set of capacitance values for RIS configuration. To obtain a low-complexity RIS control protocol, a deep reinforcement learning method, was proposed to design and update the codebook to account for time-varying channels.

% Comment: deep reinforcement learning-based method --> deep reinforcement learning method

During the transmission of uplink pilot symbols by the user, the RIS explores the updated codewords (capacitance vectors). Subsequently, the BS measures the effective channel, which dynamically changes as the RIS configuration is modified. By analyzing the measured effective channel, an optimal codeword is obtained and fed back to the RIS for configuration.

\subsection{Phase Shift Feedback from User to RIS}

For fast RIS control in FDD systems, \textbf{the RIS configuration can be received directly from the user without the participation of the BS}. Although the codebook-based method is an easy implementation for phase shift feedback from the user to the RIS, its performance is limited by the codebook size, which is much smaller than the number of RIS elements due to the limited bandwidth of the feedback channel. To feed back a large number of quantized phase shifters with small overhead, a convolutional autoencoder-based scheme was proposed in \cite{Yu-CL2022}, where the phase shift is compressed at the user and then reconstructed as the RIS. Specifically, the authors used an encoder at the user to map the RIS configuration information to a code with a smaller dimension in the feature space. The compressed code is then sent to the RIS, and a decoder at the RIS learns to recover the original configuration information from the corrupted code with noise. Building on \cite{Yu-CL2022}, the authors in \cite{Yu-arxiv} utilized the global attention mechanism to push the limit of phase shift compression by exploiting the attention map over both spatial and channel dimensions. Moreover, a simplified phase shift compression network was proposed to simplify the architecture of the decoder at the RIS.

%%%%%%%%%%%%%%%%%%%%%%%%%%%%%%%%%%%%%%%%%%%%%%%%%%%%%%%%%%%%%%%%%%%%%%%%%%%%%%%%%%%%%%%%%%%%%%

%%%%%%%%%%%%%%%%%%%%%%%%%%%%%%%%%%%%%%%%%%%%%%%%%%%%%%%%%%%%%%%%%%%%%%%%%%%%%%%%%%%%%%%%%%%%%%
\section{Future Directions}\label{sec:5}

In previous sections, we reviewed advancements in limited feedback scheme design and extracted underlying design features introduced by the RIS.  Given the growth of RIS applications, evolving architectures, and emerging wireless technologies, we highlight potential research directions in this design realm.

\subsection{RIS--user Association}
The feedback problem is much more challenging in the MUMR scenario due to the extremely rich scattering environment that is contributed by distributed RISs. Although the total BS--RISs channel is still shared by different users, the utility of each BS-RIS subchannel varies for different users because the path loss for the BS--RIS--user link can be largely different. In this context, it is possible to apply the channel customization methodology to reshape a sparse cascaded channel with respective enhanced paths for each user, thereby decoupling the complex feedback in MUMR scenarios into several easier feedback problems in SUMR scenarios. \textcolor{black}{The key to achieve this goal lies in how to design the RIS--user association with dedicated reflection designs, based on tunable link quality and spatial distribution characteristics of users and RISs, to ensure approximately orthogonal end-to-end channels for different users.}

\subsection{Near-field Feedback}
The low-cost hardware makes it possible to assemble a large number of elements on the RIS. With the increment of the RIS size, near-filed communications can occur between the RIS and the BS/user. In the near-field, parameters, such as distances between every element pair and the orientation of the array, should be considered to model the channel, which makes the channel more complicated than that in the far-field. Because existing feedback schemes developed for the far-field cannot adequately match the characteristics of near-field channels, applying these schemes to the near-filed would result in severe performance degradation. Therefore, it is necessary to develop new feedback strategies and codebooks to capture the near-field characteristics.

\subsection{Active RIS-assisted Feedback}
The passive RIS, which lacks baseband signal processing capabilities, poses significant challenges for channel estimation and feedback. A novel RIS architecture, namely active RIS, based on sparse channel sensors was first proposed in \cite{Active-RIS} to enable channel estimation at the RIS. However, obtaining the accurate CSI at the RIS still requires feedback because the estimation capability of the active RIS is limited by the small number of channel sensors. In this case, it is worth investigating how to complete channel acquisition process with limited feedback and limited estimation.

\subsection{Multi-modal Information Utilization}
CSI serves as the indicator of the propagation environment, which can also be described by other modal information, such as images and point clouds. During the CSI feedback of RIS-assisted systems, multi-modal information can be utilized to help improve feedback accuracy or reduce feedback overhead. However, it is difficult for conventional algorithms to mix the information from different modals, such as CSI and images. Artificial intelligence (AI) can be utilized to introduce other modal information to the CSI feedback process \cite{guo-arxiv}.

\subsection{AI-based Feedback}
Although AI has shown considerable potential in the feedback of RIS systems, several challenges must be addressed. First, only simulated data has been utilized thus far, and high-quality CSI samples need to be measured and collected to train and evaluate AI models. Additionally, the generalization of AI-based feedback needs to be evaluated and improved. Finally, the feedback overhead is still significant when compared to existing massive MIMO systems. Task-oriented feedback, such as the approach presented in \cite{Guo-tvt}, is a potential solution, but it requires a more thorough design.

\section{Conclusions}\label{sec:6} 
This article explores two primary use cases of RIS-associated limited feedback: channel reconstruction and RIS configuration. By delving into the feedback characteristics and associated challenges for these use cases, we elucidate the methods to exploit the channel properties presented by the RIS when formulating feedback schemes and corresponding codebooks. Given that limited feedback in RIS-assisted systems remains in its early stages, we pinpoint several prospective research areas, encompassing the MUMR scenario, near-field considerations, active RIS, the utilization of multi-modal information, and the incorporation of AI techniques.

%\section*{ACKNOWLEDGEMENT}
%\label{ACKNOWLEDGEMENT}

%This work was supported in part by National Natural Science Foundation of China (NSFC) under Grant 62401137, \textcolor{red}{62401640, and XX}, in part by the Natural Science Foundation of Jiangsu Province under Grant BK20241281, in part by the China National Postdoctoral Program for Innovative Talents under Grant BX20230065 and 2024M750421, in part by the Jiangsu Excellent Postdoctoral Program under Grant 2023ZB476, in part by the Guangdong Basic and Applied Basic Research Foundation under Grant 2023A1515110732.

%\theendnotes

\bibliographystyle{gbt7714-numerical}
\bibliography{myref}

\end{document}